\documentstyle[12pt]{article}

\newcommand{\ba}{\begin{eqnarray}}
\newcommand{\ea}{\end{eqnarray}}
\newcommand{\tr}{\,\mbox{tr}\,}

\begin{document}
\begin{titlepage}
\title{Mean field expansion and meson effects in chiral
condensate of analytically regularized Nambu-Jona-Lasinio model}

\author{
R.G. Jafarov\footnote{Physical Department of Baku State
University, Baku, Azerbaijan}, V.E. Rochev\footnote{Institute for
High Energy Physics, Protvino, Moscow region, Russia (e-mail:
rochev@mx.ihep.su)}}
\date{}
\end{titlepage}
\maketitle

\abstract{ Scalar meson contributions in chiral quark condensate
are calculated in the  analytically regularized Nambu --
Jona-Lasinio model using the framework of mean-field expansion in
bilocal-source formalism. The sigma-meson contribution for
physical values of the parameters is found to be small. Pion
contribution is found to be significant and should be taken into
account for the choice of the parameter values.}

\section{Introduction}
Nambu--Jona-Lasinio (NJL) model \cite{NJL}  was the first field
theoretical model of dynamical chiral symmetry breaking (DCSB) in
hadron physics. The NJL model has been reformulated in quark
language \cite{QNJL} during the seventies and eighties, and since
then it exists as a successful effective model of quantum
chromodynamics
 of light hadrons in the non-perturbative region. Subsequently,
 the NJL model was intensively investigated also as a model
of hadron matter at finite temperature and density \cite{TNJL}.
(For more references see also reviews \cite{kle} and
\cite{HatKun}.)

In overwhelming majority of these investigations, the NJL model
has been considered in the mean-field approximation (Hartree
approximation), or in  leading order of $1/n_c$-expansion ( $n_c$
is the number of colors). The successes in the phenomenological
applications have led to an analysis of the structure of the NJL
model in next-to-leading order of the  expansions ( see
\cite{Domi}-\cite{Ripka} and refs. therein). Such an analysis is
necessary for clarification of the region of applicability of
results and the stability under a variation of parameters and
quantum fluctuations due to higher-order effects.

Since the NJL model in the mean-field approximation includes quark
loops, the essential aspect of application of this model is a
regularization. Most common regularizations for NJL model
traditionally entail a four-dimensional cutoff in Euclidean
momenta or a three-dimensional momentum cutoff.  Other
regularization schemes (Pauli--Villars regularization or non-local
Gauss formfactors) also are used for the NJL model. The least
common regularization for NJL model is a dimensional
regularization (Thus in  reviews cited above, \cite{kle} and
\cite{HatKun} this regularization is not even mentioned). At first
it might appear somewhat strange, since the merits of dimensional
regularization are generally known, and the  same regularization
most commonly used for calculations in renormalized theories,
particularly in gauge theories.
  Apparently this fact is connected with the following circumstance:
   in contrast to renormalized models, a parameter of
   regularization in the NJL model is included in formulae for physical
   quantities,
   and it is one of the essential parameters of the model. But the
   parameter of dimensional regularization, which is traditionally
   treated
   as a deviation in physical dimension of  space, does not
permit any physical interpretation in this treatment.

However, an alternative treatment of dimensional regularization
exists -- as a variant of an analytical regularization. In this
treatment all calculations are made in four-dimensional Euclidean
momentum space, and the regularization parameter is treated as a
power of a weight function, which regularizes divergent integrals.
Such treatment of dimensional regularization, based on ideas of
Wilson and Collins, is consistently developed and applied to the
NJL model by Krewald and Nakayama \cite{Krewald} in the mean-field
approximation. It should be stressed that in this treatment of
dimensional regularization, the regularization parameter is not at
all a deviation in the physical dimension of space.

We suppose that a possible treatment of this parameter is a power
of some factor, which is a measure  of gluon influence on the
effective four-fermion quark self-action of NJL model. In some
sense this possible treatment has something in common with
non-local variants of the NJL model (see \cite{Ripka},
\cite{Volkov}).

In this work we study the NJL model with dimensional
regularization in the treatment of Krewald and
Nakayama\footnote{To avoid unnecessary associations with the usual
treatment of dimensional regularization we shall in what follows,
 refer to this regularization as analytical regularization of NJL
model} in the next-to-leading order of the mean-field expansion.
To formulate the mean-field expansion (Section 1) we have used an
iteration scheme of solution of Schwinger-Dyson equation with
fermion bilocal source, which has been developed in the literature
\cite{Ro1}. Analytical regularization of the NJL model is
discussed in Section 2. A purpose of our calculation is to study
meson contributions to quark chiral condensate, which is a
principal order parameter in models of DCSB. As our calculations
demonstrate (Section 3), the pion contribution to chiral
condensate is expressed in the analytical regularization by a very
simple formula (\ref{rp})-- a ratio of pion contribution to
leading approximation condensate is inversely proportional to the
regularization parameter and does not depend on other parameters
of the model. The pion contribution is rather significant, and
tends to infinity at a bound of admissible values of the parameter
, i.e., the model is unstable with respect to quantum fluctuations
near this bound. The sigma-meson contribution is small for
admissible values of the parameter. In Section 4 the results of
Sections 2 and 3, which were obtained for the classical variant of
the NJL model with U(1)-symmetry, are generalized for a physically
interesting model with two flavors and $n_c$ colors (SU(2)-model).

 A choice of physical values for SU(2)-model
parameters is discussed in Section 5.  A discussion of the results
is presented in the Conclusion. Quite briefly our result can be
stated as  follows: for physical values of parameters, the
analytically regularized NJL model does not contain  any
pathological quantum fluctuations, connected with the scalar-meson
contributions, though the pion contribution is significant and
should be taken into account  in phenomenological treatments of
NJL-type models.

\section{Mean-Field Expansion in Bilocal-Source Approach}

By the U(1)-model, we mean a theory of self interacting spinor
field $\psi$ defined by the Lagrangian
\begin{equation}
{\cal L} =  \bar\psi
i\hat\partial\psi+\frac{g}{2}\biggl((\bar\psi\psi)^2+(\bar\psi
i\gamma_5\psi)^2\biggr). \label{LU1}
\end{equation}
Here $g>0$ is a coupling. This Lagrangian is invariant under
transformations of the chiral group $U_V(1)\times U_A(1)$. Such a
model has no direct physical applications, but, as is shown in
Section 4, the results of physical   SU(2)-model for meson
contributions in chiral condensate are practically identical to
the results of the U(1)-model. This is not surprising since these
contributions are purely dynamical. Distinctions are displayed as
simple coefficients (see Eq. (\ref{rspSU2}) of Section 4).

A generating functional of Green functions (vacuum expectation
values of $T$-products of fields) can be represented as the
functional integral with bilocal source

\begin{equation}
G(\eta)  =\int D(\psi,\bar\psi)\exp i\Big\{\int dx{\cal L} -\int
dx dy \bar\psi(y)\eta(y,x)\psi(x)\Big\}, \label{G}
\end{equation}

where $\eta(y,x)$ is the bilocal source of the spinor field.

The functional derivative of  $G$ over source  $\eta$ is a
one-particle (two-point) Green function  (a propagator of the
field $\psi$):
\begin{equation}
\frac{\delta G}{\delta\eta(y,x)}\bigg\vert_{\eta=0} = i<0\mid
T\Big\{\psi(x)\bar\psi(y)\Big\}\mid 0>\equiv S(x-y). \label{S}
\end{equation}

The $n$-th functional derivative of  $G$ over source  $\eta$ is
the $n$-particle  ($2n$-point) Green function:
$$
 \frac{\delta^n
G}{\delta\eta(y_1,x_1)\cdots\delta\eta(y_n,x_n)}\bigg\vert_{\eta=0}
= i^n<0\mid
T\Big\{\psi(x_1)\bar\psi(y_1)\cdots\psi(x_n)\bar\psi(y_n)\Big\}\mid
0>\equiv S_n\left( \begin{array}{cc}
x_1&y_1\\\cdots&\cdots\\x_n&y_n\end{array} \right). \label{VEV}
$$
Translational invariance of the functional-integration measure in
Eq. (\ref{G}) give us the functional-differential Schwinger-Dyson
equation (SDE) for the generating functional $G$:
\begin{equation}
\delta(x-y)G + i\hat\partial_x\frac{\delta G}{\delta\eta(y,x)}
+ig\Big\{\frac{\delta}{\delta\eta(y,x)} \tr \frac{\delta
G}{\delta\eta(x,x)}-
\gamma_5\frac{\delta}{\delta\eta(y,x)}\tr\gamma_5\frac{\delta
G}{\delta\eta(x,x)}\Big\} = \label{SDE}
\end{equation}
$$
=\int dx_1\eta(x,x_1) \frac{\delta G}{\delta\eta(y,x_1)}. $$ We
shall solve this equation employing the method
 proposed in  \cite{Ro1}.
(See also the
 brief review  in  \cite{Ro2}.)

For the NJL model this method is a variant of the mean-field
expansion.

A leading approximation is an approximation of the
functional-differential SDE
 (\ref{SDE}) by the equation with zero r.h.s:
\begin{equation}
\delta(x-y)G^{(0)} + i\hat\partial_x \frac{\delta
G^{(0)}}{\delta\eta(y,x)}
+ig\Big\{\frac{\delta}{\delta\eta(y,x)}\tr\frac{\delta
G^{(0)}}{\delta\eta(x,x)}-
\gamma_5\frac{\delta}{\delta\eta(y,x)}\tr\gamma_5\frac{\delta
G^{(0)}}{\delta\eta(x,x)}\Big\} = 0 \label{G0}
\end{equation}
A solution of  Eq.(\ref{G0}) is the functional
\begin{equation}
G^{(0)}= \exp\Big\{\mbox{Tr}\,\Big(S^{(0)}\ast\eta\Big)\Big\}
\label{G^0}
\end{equation}
(here $\mbox{Tr}$ is a trace in operator sense, and  $\ast$ is
operator multiplication), where $S^{(0)}$ is a solution of
equation
\begin{equation}
\delta(x)+i\hat\partial S^{(0)}(x)  +ig\Big\{ S^{(0)}(x)\tr
S^{(0)}(0)-\gamma_5S^{(0)}(x)\tr \gamma_5S^{(0)}(0)\Big\}=0.
\label{S0}
\end{equation}
Leading approximation
 (\ref{G0})--(\ref{G^0}) generates the linear iteration scheme
$$
G = G^{(0)} + G^{(1)} + \cdots + G^{(n)} + \cdots,
$$
where $n$-th step functional
 $G^{(n)}$ is a solution of the equation
\begin{equation}
G^{(n)} + i\hat\partial \frac{\delta G^{(n)}}{\delta\eta}
+ig\Big\{\frac{\delta}{\delta\eta}\tr\frac{\delta
G^{(n)}}{\delta\eta}-
\gamma_5\frac{\delta}{\delta\eta}\tr\gamma_5\frac{\delta
G^{(n)}}{\delta\eta}\Big\} =\eta\ast \frac{\delta
G^{(n-1)}}{\delta\eta}. \label{Gn}
\end{equation}
A solution of Eq.(\ref{Gn}) is the functional
$$
G^{(n)}= P^{(n)}G^{(0)},
$$
where $P^{(n)}$ is a polynomial of  $2n$-th degree on source
$\eta$.

It  follows from (\ref{G^0}), that the unique connected Green
function of the leading approximation is the  propagator
$S^{(0)}$. Other connected Green functions  appear in the
following iteration steps.

A solution of equation
 (\ref{S0}) is  the free propagator
$$
S^{(0)}(p)= \frac{1}{m-\hat p}
$$
with dynamical mass $m$, which is a solution of the gap equation
of the NJL model:
\begin{equation}
m=-4igm\int\frac{d\tilde p}{m^2-p^2}. \label{gap}
\end{equation}
Here and below  $d\tilde p\equiv d^4p/(2\pi)^4$.

The divergent integral in the r.h.s. of equation (\ref{gap})
should be dealt with by some regularization. The chiral-symmetric,
trivial solution $m=0$, always exists. Physically preferable
(energetically favorable) solution is a solution with $m\neq 0$,
which corresponds to DCSB. Below we shall consider just such a
solution.

The first iteration  is the functional
$$ G^{(1)}=
\biggl\{\frac{1}{2}\mbox{Tr}\Big( S^{(1)}_2\ast\eta^2\Big) +
\mbox{Tr}\Big(S^{(1)}\ast\eta\Big)\biggr\}G^{(0)}.
$$
Taking into account Eqs.(\ref{G0})--(\ref{Gn}), we obtain the
first iteration equation for the
 two-particle function
\begin{equation}
S_2^{(1)}\left( \begin{array}{cc} x&y\\x'&y'\end{array} \right)= -
S^{(0)}(x-y') S^{(0)}(x'-y)+\label{S2}
\end{equation}
$$
 +ig\int dx_1\Big\{(S^{(0)}(x-x_1)
S^{(0)}(x_1-y)) \tr S_2^{(1)}\left( \begin{array}{cc}
x_1&x_1\\x'&y'\end{array} \right)-
$$
$$
 - (S^{(0)}(x-x_1)\gamma_5
S^{(0)}(x_1-y)) \tr \gamma_5S_2^{(1)}\left( \begin{array}{cc}
x_1&x_1\\x'&y'\end{array} \right)\Big\}
$$
and the equation for the first iteration corrections to the
propagator
\begin{equation}
S^{(1)}(x-y)= ig\int dx_1S^{(0)}(x-x_1) \Big\{S_2^{(1)}\left(
\begin{array}{cc} x_1&y\\x_1&x_1\end{array} \right)   -
\gamma_5 S_2^{(1)}\left( \begin{array}{cc}
x_1&y\\x_1&x_1\end{array} \right)\gamma_5\Big\}+ \label{S1}
\end{equation}
$$
 + ig\int dx_1 S^{(0)}(x-x_1)
S^{(0)}(x_1-y)\tr S^{(1)}(0).
$$
A solution of the linear integral equation  (\ref{S2}) is
$$
S_2^{(1)}\left( \begin{array}{cc} x&y\\x'&y'\end{array} \right)=
$$
\begin{equation}
=\int dx_1dy_1dx'_1dy'_1S^{(0)}(x-x_1) S^{(0)}(x'-x'_1) F_2\left(
\begin{array}{cc} x_1&y_1\\x'_1&y'_1\end{array} \right)
S^{(0)}(y_1-y) S^{(0)}(y'_1-y'), \label{S2S}
\end{equation}
where the amputated two-particle function $F_2$ is
\begin{equation}
F_2\left( \begin{array}{cc} x&y\\x'&y'\end{array} \right) =
-[S^{(0)}]^{-1}(x-y') [S^{(0)}]^{-1}(x'-y)+ \label{F2}
\end{equation}
$$
+ \delta(x-y)\delta(x'-y')\big\{1\otimes 1\cdot A_\sigma(x-x') +
\gamma_5\otimes\gamma_5\cdot A_\pi(x-x')\Big\},
$$
and scalar amplitudes $A_\sigma$  and  $A_\pi$ in momentum space
are
\begin{equation}
A_\sigma(p)=-\frac{ig}{1-L_S(p)}, \;\;\;\; L_S(p)=ig\int d\tilde q
\; \tr S^{(0)}(p+q)S^{(0)}(q),\label{Asigma}
\end{equation}
    and
\begin{equation}
A_\pi(p)=\frac{ig}{1+L_P(p)},\;\;\;\;L_P(p)=ig\int d\tilde q \;
\tr
S^{(0)}(p+q)\gamma_5S^{(0)}(q)\gamma_5. \label{Api}
\end{equation}

Taking into account the gap equation (\ref{gap}) with $m\neq 0$,
one can obtain (in translational-invariant regularization) for
$A_\sigma$ and $A_\pi$ the following representations:
\begin{equation}
A_\sigma(p)=\frac{1}{2(4m^2-p^2)I_0(p)} \label{A_sigma}
\end{equation}
\begin{equation}
 A_\pi(p)=\frac{1}{2p^2I_0(p)}.
\label{A_pi}
\end{equation}
Here
\begin{equation}
 I_0(p )=\int d\tilde
q\frac{1}{(m^2-(p+q)^2)(m^2-q^2)}. \label{I0}
\end{equation}

Eq.(\ref{S1}) for $S^{(1)}$,  taking into account the results for
$S_2^{(1)}$,  reduces in the momentum space to a system of simple
algebraic equations. Introducing the first iteration mass operator
$\Sigma^{(1)}=[S^{(0)}]^{-1}\star S^{(1)}\star [S^{(0)}]^{-1}$ we
obtain from equation (\ref{S1}):
\begin{equation}
\Sigma^{(1)}(x) =ig\delta(x)\tr S^{(1)}(0)+
S^{(0)}(x)A_\sigma(x)+S^{(0)}(-x)A_\pi(x). \label{Sigma}
\end{equation}

Let us briefly discuss the following step of the iteration scheme.
A solution of the second iteration functional-derivative equation
has the form
$$ G^{(2)}=
\biggl\{\frac{1}{4!}\mbox{Tr}\Big( S^{(2)}_4\ast\eta^4\Big)+
\frac{1}{3!}\mbox{Tr}\Big( S^{(2)}_3\ast\eta^3\Big) +
\frac{1}{2}\mbox{Tr}\Big( S^{(2)}_2\ast\eta^2\Big) + \mbox{
Tr}\Big(S^{(2)}\ast\eta\Big)\biggr\}G^{(0)},
$$
i.e., in the second iteration equations a four-particle and
three-particle functions  ($S^{(2)}_4$ and $S^{(2)}_3$) appear.
Equations for
 $S^{(2)}_2$ and $S^{(2)}$ have the same form as for the first
 iteration,
 except for  inhomogeneous terms, which contain $S^{(2)}_4$ and
$S^{(2)}_3$ for the second iteration equations.

\section{Dimensional regularization in the NJL model}

Due to non-renormalizability of the NJL model, the regularization
is an essential component of the model.

We shall use in this work the dimensional regularization in a
variant form, which was proposed in
 \cite{Krewald}. The general principles of the approach of
  \cite{Krewald}
  are the
following:
\begin{itemize}
\item All calculations are made in 4-dimensional Euclidean space;
\item Translational invariance is assumed; \item The
regularization procedure consists in transformation of integration
measure by a weight function which provides a convergence of the
integrals.
\end{itemize}
In this approach the dimensional regularization is, in essence, a
variant of analytical regularization. This point is very important
for the use and interpretation of this regularization. In this
connection we shall, in what follows, use the term "analytical"
for this regularization to stress its peculiarities in contrast to
the usual treatment of the dimensional regularization in a formal
 D-dimensional space. Let us consider the  gap equation (\ref{gap})
 of the NJL model as
an example.

The gap equation with $m\neq 0$ in Euclidean space after angle
integration is
$$
1=2g\frac{\Omega_4}{(2\pi)^4}\int \frac{q^2_e dq^2_e}{m^2+q^2_e},
$$
where $\Omega_4=2\pi^2$ is a surface of 4-dimensional unit sphere.
In accordance with  the previous comment, we modify the integrand
by the weight function:
$$
w_{\Lambda,D}(q^2_e)=w_\Lambda(q^2_e) w_D(q^2_e)
=\theta(\Lambda^2-q^2_e)\Biggr(\frac{\mu^2}{q^2_e}\Biggl)^{2-D/2}.
$$
The weight function $w_{\Lambda,D}$ is the product  of two weight
functions $w_{\Lambda}$ and $w_{D}$. The function $w_{\Lambda}$
corresponds to the four dimensional cutoff regularization, while
the function $w_{D}$ corresponds to dimensional (analytical)
regularization.

A calculation of the integral over $dq^2_e$ results in
$$
1
=\frac{2gm^2\Omega_4}{(2\pi)^4}\Biggr(\frac{m^2}{\mu^2}\Biggl)^{D/2-
2} B_{\frac{\Lambda^2}{m^2+\Lambda^2}} (D/2, 1-D/2).
$$
Here $B_x(u, v)$ is the incomplete Beta-function.

\vspace{.2in}

 a) Cutoff: By setting $D=4$,  we have
$$
1=\kappa_\Lambda\biggr(1-\frac{m^2}{\Lambda^2}
\log(1+\frac{\Lambda^2}{m^2})\biggl),
$$
where $\kappa_\Lambda=g\Lambda^2/4\pi^2$.

This relation coincides exactly  with the classical result of
\cite{NJL}.

\vspace{.2in}

b) Analytical (dimensional) regularization:  When
$\Lambda^2\rightarrow\infty$,  taking into account the formula
$$
B_1 (D/2, 1-D/2)=\Gamma(D/2)\Gamma(1-D/2)
$$
and rescaling the scale parameter $\mu^2$ as
\begin{equation}
(\mu^2)^{2-D/2}=\frac{\Omega_D}{\Omega_4}\frac{(2\pi)^4}{(2\pi)^D}
(M)^{2-D/2} \label{rescaling}
\end{equation}
we obtain for the gap equation
\begin{equation}
1=\kappa\Gamma(1-D/2)\Biggr(\frac{m^2}{4\pi M^2}\Biggl)^{D/2-2} .
\label{gapD}
\end{equation}
 Here instead of $g$ we introduce the dimensionless quantity
\begin{equation}
 \kappa=\frac{gm^2}{4\pi^2}.
 \label{kappa}
 \end{equation}
Equation  (\ref{gapD})  corresponds exactly to the calculation of
the initial integral with the formal prescription of D-dimensional
integration
$$
d\tilde q \equiv \frac{d^4q}{(2\pi)^4}\rightarrow
\frac{(M^2)^{2-D/2}d^Dq}{(2\pi)^D},
$$
but in our case  the calculation was performed in the usual
4-dimensional space, i.e. in our treatment $D$ is not a dimension
of some space, but a parameter which  provides the convergence. In
particular, we are not constrained by the limit $D\rightarrow 4$
for the analysis of the results. We suppose that a possible
treatment of regularization parameter is a power of some
additional  factor, which is a measure of gluon influence on the
effective local four-quark self-interaction of the NJL model.

Below we shall use the regularization parameter $\xi$ as
\footnote{Note that parameter $\xi$  is different from the
commonly used parameter $\varepsilon=(4-D)/2.$ They are connected
by the relation $\varepsilon=1+\xi$. Introduction of this notation
prevents unnecessary associations with the usual treatment of
dimensional regularization. Furthermore, in terms of the parameter
$\xi$ all subsequent formulae of the NJL model acquire simple
forms.}
\begin{equation}
D=2-2\xi. \label{xi0}
\end{equation}

In terms of the parameter  $\xi$ gap equation (\ref{gapD}) has the
form
\begin{equation}
1=\kappa\Gamma(\xi)\Biggr(\frac{4\pi M^2} {m^2}\Biggl)^{1+\xi} .
\label{gapxi}
\end{equation}

The region of  convergence of the integral is $ 0<\xi<1.$ As we
shall see (Section 5), this is also the region for the physical
values of the model parameters.

Chiral condensate in the leading approximation is
$$
(c^3)^{(0)}=i\tr S^{(0)}(0)=-\frac{m^3}{4\pi^2}\Biggr(\frac{4\pi
M^2} {m^2}\Biggl)^{1+\xi} \Gamma(\xi)=-\frac{m}{g}.
$$

Integral $I_0$ (see (\ref{I0})), which is a part of scalar
amplitudes $A_\sigma$ and  $A_\pi$,  can also be calculated as
above. Transforming to Euclidean metric, introducing a standard
Feynman parameterization, and changing an integration variable
(which is possible due to translational invariance of the
procedure, see \cite{Krewald}) we can perform the angular
integration. According to the procedure,  we introduce into the
integrand, a weight function $w_D(q_e^2)$  and, after the same
rescaling (\ref{rescaling}), obtain the result, which also exactly
corresponds to the result of integration with the formal
transition to $D$-dimensional space.

Taking into account the gap equation (\ref{gapxi}), we obtain the
pole approximation for the scalar amplitudes in analytical
regularization
\begin{equation}
A_\sigma(p)\simeq\frac{1}{2(4m^2-p^2)I_0(4m^2)}=
\frac{2igm^2(1+2\xi)}{(4m^2-p^2)\xi}, \label{Aspole}
\end{equation}
\begin{equation}
A_\pi(p)\simeq\frac{1}{2p^2I_0(0)}= -\frac{2igm^2}{p^2\xi},
\label{Appole}
\end{equation}
which correspond to contributions of sigma-meson and pion. We
shall use these expressions in the following calculations.

\section{Meson contributions to the chiral condensate}

As a measure of quantum fluctuations of the chiral field, consider
a ratio of first iteration condensate
\begin{equation}
(c^3)^{(1)}=i\mbox{tr}S^{(1)}(0) \label{xi1}
\end{equation}
to the leading-approximation condensate   $(c^3)^{(0)}$:
\begin{equation}
r\equiv \frac{(c^3)^{(1)}}{(c^3)^{(0)}}=r_\sigma+r_\pi. \label{r}
\end{equation}
Here $r_\sigma$ is a  contribution of the scalar meson
(sigma-meson) and $r_\pi$ is a  contribution of the pseudoscalar
meson (pion).

  From Eq.(\ref{Sigma}), we obtain
$$
r_\sigma=-\frac{2ig}{\xi}\int \frac{d\tilde p d\tilde q
(m^2+3p^2-2(pq))}{(m^2-p^2)^2(m^2-(p-q)^2)} A_\sigma(q),
$$
and
$$
r_\pi=-\frac{2ig}{\xi}\int \frac{d\tilde p d\tilde q
(m^2-p^2+2(pq))}{(m^2-p^2)^2(m^2-(p-q)^2)} A_\pi(q).
$$
Calculation of these two-loop integrals gives us the desired
answer.

The integrals for $r_\pi$ are calculated in the analytical
regularization in closed form and give us a very simple expression
\begin{equation}
r_\pi=\frac{1}{4\xi}. \label{rp}
\end{equation}

The scalar contribution can be represented as
\begin{equation}
r_\sigma=\frac{4^\xi\Gamma(\frac{3}{2}+\xi)}
{2\sqrt{\pi}\Gamma(3+\xi)} \int_0^1\frac{du}{(4-3u)^{1+\xi}}
\biggl[(3-2u)(1-\xi) F(1+\xi,2-\xi; 3+\xi; \frac{(u-2)^2}{4-3u})-
\label{rs}
\end{equation}
$$
-(1+2\xi)F(1+\xi,1-\xi; 3+\xi; \frac{(u-2)^2}{4-3u})\biggr].
$$
where $F(a, b; c; z)$ is the Gauss hypergeometric function.

Note, that $r_\pi$ and $r_\sigma$ depend only on the parameter
$\xi$  and do not depend on other parameters of the model. This
feature is a peculiarity of the analytical regularization.

Results of the calculations in the region $0<\xi\le 1$ are shown
in Table 1. One can see that the contribution of the sigma-meson
is small, while the contribution of the pion in the region
$0<\xi\le 0.3\div0.5$ is significant and, strictly speaking, in
this last region we cannot consider this contribution as a mere
correction, i.e., at such values of the parameter $\xi$ the
quantum fluctuations are large and can lead to a principal
modification of the entire physical picture of the NJL model (in
spirit of work \cite{Kleinert}, for example).

\begin{table}[!ht]
\begin{center}
\begin{tabular}{l|ccc}\hline
 ${\xi}$ & $r_{\sigma}$ & $r_{\pi}$ & $c^{(1)}/c^{(0)}$\\
 \hline\hline
\  0.1  & \ 0.264  & \  2.50  & \   0.556 \\
\  0.2  & \ 0.189 &  \  1.250 & \   0.346 \\
\  0.3  & \ 0.119 &  \  0.833 & \   0.250 \\
\  0.4 &  \ 0.057 &  \  0.625 & \   0.189 \\
\  0.5 &  \ 0     &  \  0.50  & \   0.145 \\
\  0.6 & -0.050 &    \  0.417 & \   0.110 \\
\  0.7 & -0.094 &    \  0.357 & \   0.081 \\
\  0.8 & -0.131 &    \  0.313 & \   0.057 \\
\  0.9 & -0.162 &    \  0.278 & \   0.037 \\
\  1.0 &   -0.188 &    \  0.250 & \   0.021 \\
 \hline
\end{tabular}
\caption{Relative contributions of sigma-meson
($r_\sigma$) and pion ($r_\pi$) to the chiral
condensate $c^3$ of U(1)-model and a ratio ($c^{(1)}/c^{(0)}$)  of
first iteration condensate to leading-approximation
 condensate as  functions of the regularization parameter $\xi$.}
\end{center}
\end{table}

\section{SU(2)-model}

The U(1)-model has no direct physical applications, and
 we
cannot  estimate  the physical values of model parameters in the
framework of this model and, thus cannot  reach any definite
conclusion about the  meson contributions.

To study the possible values of the parameters $m$, $\kappa$ and
$\xi$, let us consider a model connected with light hadron
phenomenology. Accordingly we study the model defined by the
Lagrangian
\begin{equation}
{\cal L}=\bar \psi i\hat \partial\psi+\frac{g}{2}
\biggl[(\bar\psi\psi)^2+(\bar\psi i\gamma_5 \tau^a\psi)^2\biggr].
\label{LSU2}
\end{equation}
Here $ \psi\equiv\psi^{\alpha, c}_j $;\quad  $\alpha=1, 2, 3, 4$
is Dirac spinor index; $c=1,\ldots, n_c$ is color index; $j=1,2$
is isotopic (flavor) index; $\tau^a$ are generators of $SU(2)$
 (Pauli matrices); $a= 1, 2, 3$. This model has the chiral
symmetry of $SU_V(2)\times SU_A(2)$-group. We call this the
 SU(2)-model.

Mean-field expansion in bilocal-source formalism is constructed
with the same scheme as in the U(1)-model above, and so without
going to the details,  we enumerate only the major differences
from corresponding results of U(1)-model.

In the leading approximation, the propagator is diagonal over
color and flavor:
\begin{equation}
S^{(0)}_{cd, jk}= \delta_{cd}\delta_{jk}(m-\hat p)^{-1},
\label{S0SU2}
\end{equation}
and the gap equation for the SU(2)-model is
\begin{equation}
1= -8i\, g\, n_c\, \int\frac{d\tilde p}{m^2-p^2}. \label{gapSU2}
\end{equation}
The chiral condensate is defined for each flavor, i.e.,
$$
c^3_u =<0\vert\bar u u\vert 0>,\;\;c^3_d =<0\vert\bar d d\vert 0>.
$$
(In the chiral limit  $c_d=c_u$.) Two-particle first iteration
amplitude
 $A$ (connected part of amputated two-particle function $S_2^{(1)}$)
  has the following color and flavor structure:
\begin{equation}
A^{cd, jk}_{c'd', j'k'}
=\delta^{cd}\delta^{c'd'}\biggl[\delta_{jk}\delta_{j'k'}A_\sigma
 + \tau^a_{jk}\tau^a_{j'k'}
A_\pi \biggr]. \label{ASU2}
\end{equation}
Scalar amplitudes   $A_\sigma$ and $A_\pi$ are defined by formulae
 (\ref{Asigma}) and (\ref{Api}), but traces in the definitions of
  loops $L_S$ and $L_P$ are now taken over all discrete
 indices.

First iteration mass operator $\Sigma^{(1)}=[S^{(0)}]^{-1}\star
S^{(1)}\star [S^{(0)}]^{-1}$ is diagonal over color and flavor,
and is connected with the scalar amplitudes by relation
\begin{equation}
\Sigma^{(1)}(x)^{cd}_{jk}
=\delta^{cd}\delta_{jk}\cdot\biggl[ig\delta(x)\tr S^{(1)}(0)+
S^{(0)}(x)A_\sigma(x)+3S^{(0)}(-x)A_\pi(x)\biggr].
\label{SigmaSU2}
\end{equation}

For the ratio of the first iteration condensate to the
leading-approximation condensate we obtain
\begin{equation}
r= r_\sigma+r_\pi = \label{rSU2}
\end{equation}
$$
=-\frac{8\,i\, g \,n_c\, }{1-8\, i\, g \,n_c \, J}\int
\frac{d\tilde q d\tilde p}
{(m^2-p^2)^2(m^2-(p-q)^2)}\biggl((3p^2-2(pq)+m^2)A_\sigma(q)
+3(m^2-p^2+2(pq))A_\pi(q)\biggr),
$$
where
$$
J=\int d\tilde p\frac{m^2+p^2}{(m^2-p^2)^2}.
$$

The gap equation in analytical regularization for the SU(2)-model
has exactly same form  as in (\ref{gapxi}), if we redefine the
dimensionless constant $\kappa$ as
\begin{equation}
\kappa=\frac{gn_cm^2}{2\pi^2}=2n_c\kappa_0 \label{kappaSU2}
\end{equation}
(Here and henceforward,  the index 0 denotes corresponding
quantities of the
 U(1)-model.)

 The scalar amplitudes of SU(2)-model can be written as
\begin{equation}
A_\sigma=\frac{1}{4n_c(4m^2-p^2)I_0}, \label{A_sigmaSU2}
\end{equation}
\begin{equation}
A_\pi=\frac{1}{4n_cp^2I_0}, \label{A_piSU2}
\end{equation}
where $I_0$ is defined by formula (\ref{I0}).

Taking into account the above relations and definitions we obtain
for the meson contributions of the SU(2)-model:
\begin{equation}
r_\sigma=\frac{1}{2n_c}r_{0\sigma},\;\;r_\pi=\frac{3}{2n_c}r_{0\pi},
\label{rspSU2}
\end{equation}
where $r_{0\sigma}$ and $r_{0\pi}$ are corresponding contributions
of the U(1)-model.

For physical values of colors $n_c=3$,  both contributions
decrease compared to the corresponding contributions of the
U(1)-model: $r_\sigma$ to six times, and   $r_\pi$ to half.
Accordingly the bounds of the region of large fluctuations are
moved. Results of the calculations in region  $0<\xi\le 1$ are
shown in Table 2. We see, that for the SU(2)-model at  $\xi\ge
0.2$ the first iteration condensate $c^{(1)}$ is not more than
20\% different from the leading-approximation condensate, i.e., we
may consider these values of the regularization parameter to be in
the stability zone with respect to fluctuations caused by the
meson contributions.

\begin{table}[!ht]
\begin{center}
\begin{tabular}{l|ccc}\hline
 ${\xi}$ & $r_{\sigma}$ & $r_{\pi}$ & $c^{(1)}/c^{(0)}$ \\
 \hline\hline
\  0.1  & \ 0.044  & \  1.250  &     0.319\\
\  0.2  & \ 0.032 &  \  0.625 &      0.183\\
\  0.3  & \ 0.020 &  \  0.417 &      0.128\\
\  0.4 &  \ 0.010 &  \  0.313 &      0.098\\
\  0.5 &  \ 0     &  \  0.250 &      0.077\\
\  0.6 & -0.008 &    \  0.209 &      0.063\\
\  0.7 & -0.016 &    \  0.179 &      0.052\\
\  0.8 & -0.022 &    \  0.157 &      0.043\\
\  0.9 & -0.027 &    \  0.139 &      0.036\\
\  1.0 &   -0.031 &    \  0.125 &      0.030\\
 \hline
\end{tabular}
\caption{Relative contributions of sigma-meson
  ($r_\sigma$) and pion
 ($r_\pi$) to chiral condensate  $c^3$
  of SU(2)-model and  ratio of  first-step
 condensate  to leading-approximation condensate
 $(c^{(1)}/c^{(0)}$) as  functions of regularization
 parameter $\xi$.}
\label{table2}
\end{center}
\end{table}

\section{Choice of parameters}
In determining the SU(2)-model parameters (dynamical quark mass
$m$, regularization parameter  $\xi$ and coupling  $g$ (or
dimensionless constant $\kappa=gn_cm^2/2\pi^2$)), it is necessary
to connect the values of the parameters with measurable
quantities. We accordingly choose the values of the pion decay
constant, the chiral condensate and the decay width of
$\pi^0$-meson into two photons: $\pi^0\rightarrow \gamma\gamma.$

 Pion decay constant
  $f_\pi$= 93 MeV  is defined by
\begin{equation}
i\delta^{ab}k_\mu f_\pi=<0\vert\bar
\psi\gamma_\mu\gamma_5\frac{\tau^a}{2}\psi\vert b, k>, \label{fpi}
\end{equation}
where $\vert b, k>$ is a pion state  $b$ with momentum $k_\mu$.

For the SU(2)-model the following formula for $f_\pi$  (see, for
example,
 \cite{kle}):
\begin{equation}
f^2_\pi=-4in_cm^2I_0(0) \label{fNJL}
\end{equation}
 is regularization-independent.  In the analytical
regularization we have $I_0(0)=i\xi/16\pi^2\kappa$ and  obtain:
\begin{equation}
f^2_\pi=\frac{\xi}{2g}. \label{fxi}
\end{equation}
The leading-approximation chiral condensate  is
\begin{equation}
c=(<0\vert\bar\psi\psi\vert 0>/2)^{1/3}=-(m/2g)^{1/3}\, ,
 \label{chi}
 \end{equation}
which is regularization-independent. Decay width
$\Gamma_{\pi^0\gamma\gamma}$= 7.7 KeV  has been calculated in the
analytical regularization for  SU(2)-model in ref.\cite{Krewald}.
In our notation, this formula has the form
 \begin{equation}
 \Gamma_{\pi^0\gamma\gamma}=\frac{\alpha^2m^3_\pi\xi^2(1+\xi)^2}
 {64\pi^3f^2_\pi\kappa^2}.
\label{Gamma}
\end{equation}
Here $\alpha=1/137$ is the fine structure, $m_\pi$=135 MeV is  the
mass of $\pi^0$-meson.

Formulae (\ref{fxi})--(\ref{Gamma}) enable us to determine the
 SU(2)-model parameters.

Chiral condensate $c=(<\bar \psi\psi>/2)^{1/3}$ is not a directly
measurable quantity. We determine the parameter values for some
typical values of the chiral condensate.

At $c=-160$ MeV we have $\xi\cong 1,\;m\cong 475$ MeV,
$\kappa\cong 2$. Note, that for this value of regularization
parameter $\xi$, the correction to condensate $c$ is about 3\%
(see Table 2), i.e. the model is stable with respect to quantum
fluctuations caused by the mesons. On the other hand, such a low
value of the chiral condensate hardly agrees with phenomenology,
since the Gell-Mann--Oakes--Renner formula  leads to large values
of current quark masses. At  $c=-200$ MeV we obtain $\xi\cong
0.44,\;m\cong 400$ MeV,  $\kappa\cong 0.62$. The correction to
condensate  $c$ is  9\%.
 At $c=-250$ MeV, we obtain $\xi\cong 0.2,\;m\cong 370$ MeV,
  $\kappa\cong
 0.24$,
 and the correction to condensate $c$ is more then 18\%.
Thus fixing  the model parameters with formulae
 (\ref{fxi})--(\ref{Gamma}) at phenomenologically  acceptable
 condensate values
 $c=-(200\div 250)$ MeV corresponds to the condensate
 corrections of the order
 of  $10\div20$\%.

  The model parameters as chosen above, were implemented in the
 leading-approximation formula (\ref{chi}) for the chiral condensate.
The calculated first iteration correction  modifies the choice of
parameters by the following modification of the chiral-condensate
formula:
\begin{equation}
c=-\Big(\frac{m^*}{2g^*}[1+r(\xi^*)]\Big)^{1/3}.
 \label{chi*}
 \end{equation}
(Accordingly, (\ref{fxi}) and (\ref{Gamma}) are the same with the
substitution $m\rightarrow m^*,\;g\rightarrow
g^*,\;\kappa\rightarrow \kappa^*,\;\xi\rightarrow \xi^*$.)

This modified choice of  parameters gives:

At $c=-200$ MeV: $\xi^*\cong 0.56,\; m^*\cong 420$ MeV,
$\kappa^*\cong 0.86$; the  condensate correction is  7\%;

At $c=-250$ MeV: $\xi^*\cong 0.3,\; m^*\cong 380$ MeV,
$\kappa^*\cong 0.39$; the condensate correction is  13\%.

Upon comparison of the given values with the ones above, the
modification of the parameter choice with formula  (\ref{chi*})
diminishes the relative variation of the condensate, i.e.,
stabilizes the situation. It is a consequence of the positivity of
the principal (pion) correction to the leading-approximation
condensate.

\section{Conclusion}

According to the results obtained, the analytically regularized
 NJL model gives us simple closed formulae not only for the
scalar amplitudes and the pion decay constant,  but also for the
pion contribution to the chiral condensate. As it follows from our
results, in the analytically regularized  NJL model, this pion
contribution is significant and should be taken into account with
a choice of physical values of the model parameters. At the same
time, the presented analysis demonstrates that in the analytically
regularized  NJL model, for physical values of the parameters,
taking into account next-to-leading order of the mean-field
expansion, does not lead  to any pathologies such as a
disappearance of the DCSB order parameters or of the Goldstone
boson in the spirit of ref.\cite{Kleinert, Fujita}.

\section*{Acknowledgment}
Authors are grateful to Profs. S.A. Hadjiev and K.G. Klimenko for
useful discussions.

\end{document}